\documentclass[aps,prl,preprint,amsmath,amssymb,showpacs,preprintnumbers,superscriptaddress]{revtex4-2}
\usepackage{CJK}
\usepackage{graphicx}
\usepackage{dcolumn}
\usepackage{bm}
\usepackage{tabularx}
\usepackage{multirow}
\usepackage{lettrine}

\begin{document}

\title{Neutrinoless double-$\beta$ decay and double Gamow-Teller transitions}
\author{Y. K. Wang}
\affiliation{State Key Laboratory of Nuclear Physics and Technology, School of Physics, Peking University, Beijing 100871, China}
\author{P. W. Zhao}
\email{pwzhao@pku.edu.cn}
\affiliation{State Key Laboratory of Nuclear Physics and Technology, School of Physics, Peking University, Beijing 100871, China}
\author{J. Meng}
\email{mengj@pku.edu.cn}
\affiliation{State Key Laboratory of Nuclear Physics and Technology, School of Physics, Peking University, Beijing 100871, China}
\affiliation{China Institute of Atomic Energy, Beijing 102413, China}

\date{\today}
\begin{abstract}
The neutrinoless double-$\beta$ ($0\nu\beta\beta$) decay and the double Gamow-Teller (DGT) transition are investigated with the state-of-the-art Relativistic Configuration-interaction Density functional theory.
A strong linear correlation between the nuclear matrix elements (NMEs) of the $0\nu\beta\beta$ decay and the DGT transition is demonstrated.
This linear correlation is found to originate from the similarity of the leading-order term of the $0\nu\beta\beta$-decay operator and the DGT-transition one, as revealed by expanding the $0\nu\beta\beta$-decay operator in terms of the spherical harmonics.
The present results provide a strong support to constrain the $0\nu\beta\beta$-decay NMEs through the double charge-exchange reactions.
\end{abstract}
\maketitle
\date{today}
\newpage

The search for neutrinoless double-$\beta$ ($0\nu\beta\beta$) decay is one of the top priorities in the field of nuclear and particle physics and has received worldwide attentions~\cite{Aalseth2018Phys.Rev.Lett.132502,Adams2020Phys.Rev.Lett.122501,Agostini2020Phys.Rev.Lett.252502,Albert2018Phys.Rev.Lett.072701,
Armengaud2021Phys.Rev.Lett.181802,Arnold2017Phys.Rev.Lett.041801,Gando2016Phys.Rev.Lett.082503,Dai2022Phys.Rev.D032012}.
The nuclear matrix element that governing the $0\nu\beta\beta$-decay half-life is a key issue for the decay process and its evaluation is being pursued energetically by the community~\cite{Engel2017Rep.Prog.Phys.046301,Yao2022ProgressinParticleandNuclearPhysics103965,Agostini2023Rev.Mod.Phys.025002}.
However, the discrepancy of the NMEs between model predictions is as large as a factor around four, and this limits severely the capability to anticipate the reach of the future $0\nu\beta\beta$-decay experiments and the extraction of the effective neutrino mass once the decay signals are observed.

Given the difficulty of evaluating the NMEs, the double Gamow-Teller (DGT) transition, a double spin-isospin flip mode, has been proposed to shed light on the values of NMEs~\cite{Rodriguez2013PhysicsLettersB174178,Cappuzzello2015TheEuropeanPhysicalJournalA145,takahisa2017double,Santopinto2018Phys.Rev.C061601}.
A good linear correlation between the NMEs of the $0\nu\beta\beta$ decay and those governing the ground-state-to-ground-state DGT transition has been found from the calculations of the configuration-interaction shell model (CISM) and the nonrelativistic density functional theory (NRDFT)~\cite{Shimizu2018Phys.Rev.Lett.142502}.
By studying the behavior of the NMEs as a function of the relative distance between two decaying nucleons, the correlation is attributed to the dominant short-range character of both transitions.
These results open the door to constrain the $0\nu\beta\beta$-decay NMEs from the DGT transition.

The correlation between the $0\nu\beta\beta$ decay and the DGT transition is then investigated extensively within the framework of the quasiparticle random-phase approximation (QRPA)~\cite{ifmmodeSelseSfiimkovic2018Phys.Rev.C064325,Lv2023Phys.Rev.CL051304,Jokiniemi2023Phys.Rev.C044316}, the valence-space in-medium similarity renormalization group (VS-IMSRG), and the in-medium generator coordinate method (IM-GCM)~\cite{Yao2022Phys.Rev.C014315}.
Nevertheless, all these investigations predict much weaker correlations.
Moreover, the short-range dominant behavior of the DGT transition, as revealed in the CISM calculation, is not well supported in these calculations.
The sensitive dependence of the NMEs on the isoscalar pairing strength in the QRPA~\cite{ifmmodeSelseSfiimkovic2018Phys.Rev.C064325,Lv2023Phys.Rev.CL051304,Jokiniemi2023Phys.Rev.C044316} and the in-medium renormalization effect from the IMSRG evolution in the VS-IMSRG and IM-GCM calculations~\cite{Yao2022Phys.Rev.C014315} are considered to be the possible reasons to weaken the correlation.
It should be emphasized that the conclusions in the QRPA are obtained under the assumption that nuclei have spherical symmetry, while those in the VS-IMSRG and the IM-GCM are limited to the decay processes in very light nuclei.
Therefore, whether and why there exist correlations between the $0\nu\beta\beta$ decay and the DGT transition are still important open questions.
Exploring the correlation for nuclei that are relevant to the current and next-generation experiments based on advanced nuclear many-body approaches is highly desirable.

In this Letter, the correlation between the $0\nu\beta\beta$ decay and the DGT transition is investigated within the framework of \textit{Re}lativistic \textit{C}onfiguration-interaction \textit{D}ensity functional (ReCD) theory.
The ReCD theory combines the advantages of the CISM~\cite{Caurier2005Rev.Mod.Phys.427488} and the relativistic DFT~\cite{Meng2016} and allows fully microscopic and self-consistent calculations for nuclear properties within a full model space.
It has been successfully applied to describe nuclear spectroscopic properties~\cite{Zhao2016Phys.Rev.C041301,Wang2022Phys.Rev.C054311,Wang2024Phys.Lett.B138346} and nuclear $\beta\beta$ decays~\cite{Wang2023arXivpreprintarXiv2304.12009}.
Compared to the CISM, VS-IMSRG, and IM-GCM, the ReCD theory can be applied to all the $\beta\beta$-decay candidates.
Moreover, the breaking of spherical and axial symmetries, which is important for describing the $0\nu\beta\beta$ decay~\cite{Jiao2017Phys.Rev.C054310,Wang2021Phys.Rev.C014320,Wang2023arXivpreprintarXiv2304.12009}, is considered.
Based on the ReCD theory, the NMEs of the DGT transition and the $0\nu\beta\beta$ decay in nuclei $^{48}$Ca, $^{76}$Ge, $^{82}$Se, $^{96}$Zr, $^{100}$Mo, $^{116}$Cd, $^{124}$Sn, $^{128}$Te, $^{130}$Te, and $^{136}$Xe, which are most relevant to the $0\nu\beta\beta$ decay experiments, are evaluated.
A strong linear correlation between the $0\nu\beta\beta$ decay and the DGT transition is demonstrated, providing a strong support to constrain the $0\nu\beta\beta$-decay NMEs through the DGT transition experiments~\cite{Cappuzzello2015TheEuropeanPhysicalJournalA145,takahisa2017double,Santopinto2018Phys.Rev.C061601}.
Nevertheless, the present results do not support the conclusion that the linear correlation originates in the dominant short-range character of both transitions as proposed in Ref.~\cite{Shimizu2018Phys.Rev.Lett.142502}.
Instead, by expanding the $0\nu\beta\beta$-decay operator in terms of the spherical harmonics, the presence of the linear correlation is suggested to be from the similarity of the leading-order term of the $0\nu\beta\beta$-decay operator and the DGT-transition operator. 

The $0\nu\beta\beta$-decay NME can be separated into five parts,
\begin{equation}\label{eq:0vbb-NME}
  M^{0\nu} = M^{0\nu}_{VV} + M^{0\nu}_{AA} + M^{0\nu}_{AP} + M^{0\nu}_{PP} + M^{0\nu}_{MM},
\end{equation}
where VV, AA, AP, PP, and MM represent respectively the vector, axial-vector, axial-vector and pseudoscalar, pseudoscalar, and weak-magnetism coupling channels~\cite{Song2014Phys.Rev.C054309}.
The closure approximation that is reliable for the calculation of $M^{0\nu}$ is adopted to avoid the explicit calculation of the odd-odd intermediate nuclear states~\cite{Senkov2013Phys.Rev.C064312}.
Each term shown in Eq.~\eqref{eq:0vbb-NME} can be expressed as a ``sandwich" form with a decay operator between initial and final nuclear wavefunctions, i.e., $M^{0\nu}_\alpha = \langle\Psi_f|\hat{O}^{0\nu}_\alpha|\Psi_i\rangle$.
The detailed formula of $M^{0\nu}$ can be found in Ref.~\cite{Wang2023arXivpreprintarXiv2304.12009}.
The NME of the DGT transition is defined as,
\begin{equation}\label{eq:DGT-NME}
  M^{\mathrm{DGT}} = \langle \Psi_f|\sum_{n,m}[\hat{\bm{\sigma}}_n\otimes\hat{\bm{\sigma}}_m]^0\hat{\tau}^+_n\hat{\tau}^+_m|\Psi_i\rangle,
\end{equation}
where $\hat{\bm{\sigma}}$ and $\hat{\tau}^+$ are respectively the spin and isospin-raising operators.

The initial and final nuclear wavefunctions in Eqs.~\eqref{eq:0vbb-NME} and \eqref{eq:DGT-NME} are given by the ReCD theory.
They are expressed as linear combinations of the projected wavefunctions with good angular momentum $I$ and $M$,
\begin{equation}\label{eq:wavefunction}
  |\Psi^{IM}\rangle = \sum_{K\kappa} F^{I}_{K\kappa}\hat{P}^I_{MK}|\Phi_\kappa\rangle.
\end{equation}
Here, $\hat{P}^I_{MK}$ is the three-dimensional angular momentum projection operator~\cite{Ring2004}, and $|\Phi_\kappa\rangle$ represent a set of intrinsic multi-quasiparticle states determined by solving the triaxial relativistic Hartree-Bogoliubov (TRHB) equation~\cite{Meng2016}.
The set \{$|\Phi_\kappa\rangle$\} forms the intrinsic configuration space of the ReCD theory and more details about its construction can be found in Refs.~\cite{Wang2023arXivpreprintarXiv2304.12009,Wang2024Phys.Lett.B138346}.
The expansion coefficients $F^{I}_{K\kappa}$ are obtained by diagonalizing the Hamiltonian in a shell-model space spanned by the basis $\{\hat{P}^I_{MK}|\Phi_\kappa\rangle\}$, and this leads to the Hill-Wheeler equation,
\begin{equation}\label{eq:Hill-Wheeler}
  \sum_{K\kappa}\{H^I_{K'\kappa'K\kappa} - E^IN^I_{K'\kappa'K\kappa}\}F^I_{K\kappa} = 0.
\end{equation}
The energy kernel $H^I_{K'\kappa'K\kappa} = \langle\Phi_{\kappa'}|\hat{H}\hat{P}^I_{K'K}|\Phi_{\kappa}\rangle$ and the norm matrix $N^I_{K'\kappa'K\kappa} = \langle\Phi_{\kappa'}|\hat{P}^I_{K'K}|\Phi_{\kappa}\rangle$ are evaluated by the Pfaffian algorithms~\cite{Carlsson2021Phys.Rev.Lett.172501,Hu2014PhysicsLettersB162166}.

It should be emphasised that the Hamiltonian $\hat{H}$ and the TRHB equation are derived with the same density functional.
In the present work, two well-known relativistic density functionals PC-PK1~\cite{Zhao2010Phys.Rev.C054319} and PC-F1~\cite{Buervenich2002Phys.Rev.C044308} are adopted.
A finite-range separable force with strength $G = 728$ MeV fm$^3$~\cite{Tian2009PhysicsLettersB4450} is used to treat the pairing correlations.
The intrinsic states $|\Phi_\kappa\rangle$ are obtained by solving the TRHB equation in a set of three-dimensional harmonic oscillator bases in Cartesian coordinates with 10 major shells.
Similar to our previous investigations~\cite{Zhao2016Phys.Rev.C041301,Wang2022Phys.Rev.C054311,Wang2024Phys.Lett.B138346,Wang2023arXivpreprintarXiv2304.12009}, a sufficiently large intrinsic configuration space truncated by a quasiparticle excitation energy cutoff $E_{\mathrm{cut}} = 5.0$ MeV is adopted.
Our calculations are free of adjustable parameters.

\begin{figure}[htbp]
  \centering
  \includegraphics[width=1.0\textwidth]{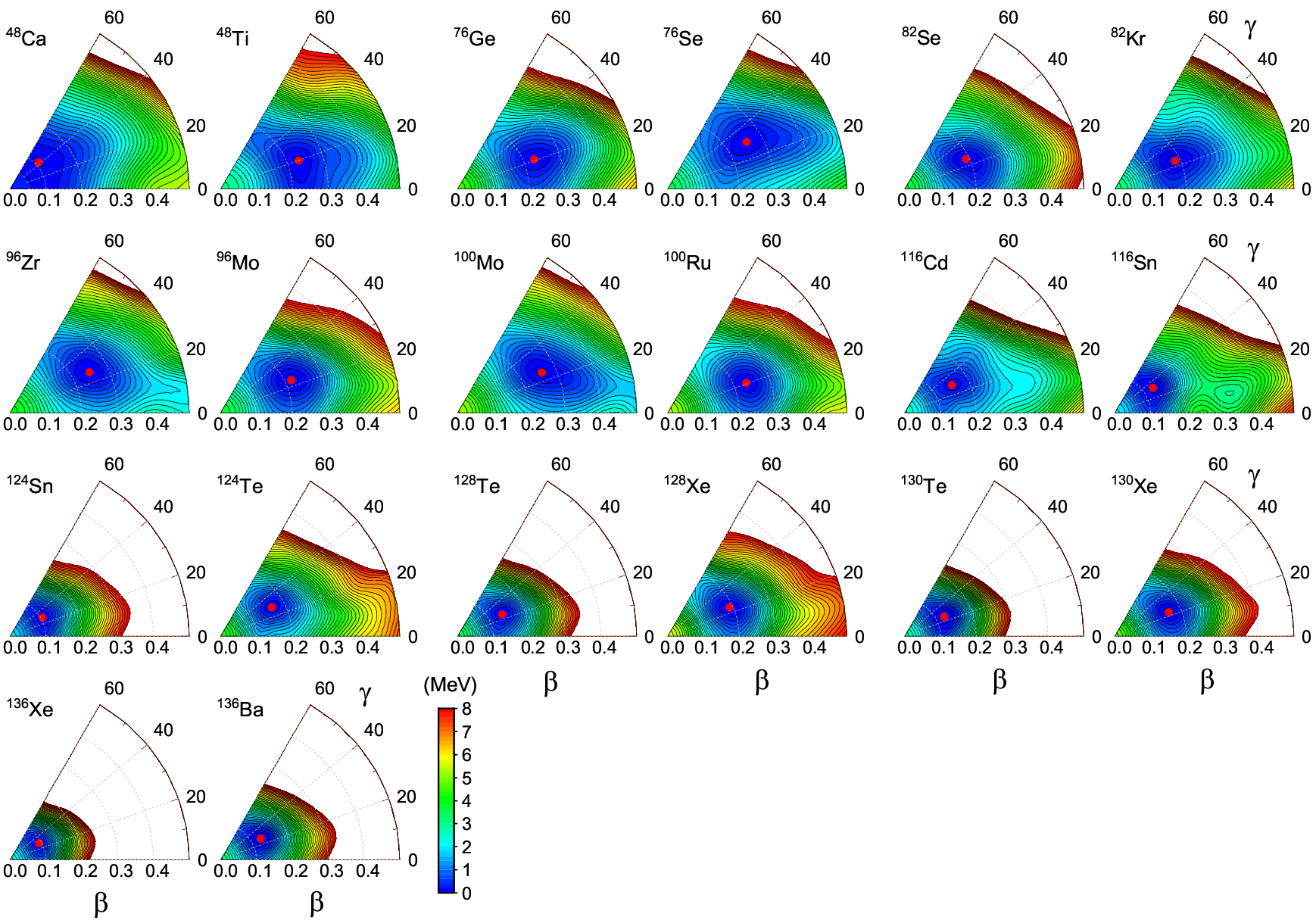}
  \caption{(Color online) Potential energy surfaces of the $0^+$ states for nuclei $^{48}$Ca, $^{76}$Ge, $^{82}$Se, $^{96}$Zr, $^{100}$Mo, $^{116}$Cd, $^{124}$Sn, $^{128}$Te, $^{130}$Te, and $^{136}$Xe, and their $\beta\beta$-decay daughter nuclei calculated by the ReCD theory with the PC-PK1 functional.
  The red solid dots denote the positions of the energy minima and the neighboring contour lines are separated by 0.2 MeV.}
  \label{Fig:PES}
\end{figure}

Figure~\ref{Fig:PES} depicts the potential energy surfaces (PESs) of the $0^+$ states calculated by the ReCD theory for ten $\beta\beta$-decay partners that are most relevant to the current $0\nu\beta\beta$ decay experiments.
The PESs based on the PC-F1 density functional are similar to those of the PC-PK1 and therefore, only the PESs obtained with the PC-PK1 are shown.
The triaxial energy minima are seen for all nuclei considered here, indicating a crucial role of the triaxiality in describing the corresponding $0\nu\beta\beta$ decays and DGT transitions.
With the wavefunctions that minimize the energies of the $0^+$ states, the NMEs for both $0\nu\beta\beta$ decay and DGT transition are evaluated.

\begin{figure}[htbp]
  \centering
  \includegraphics[width=0.5\textwidth]{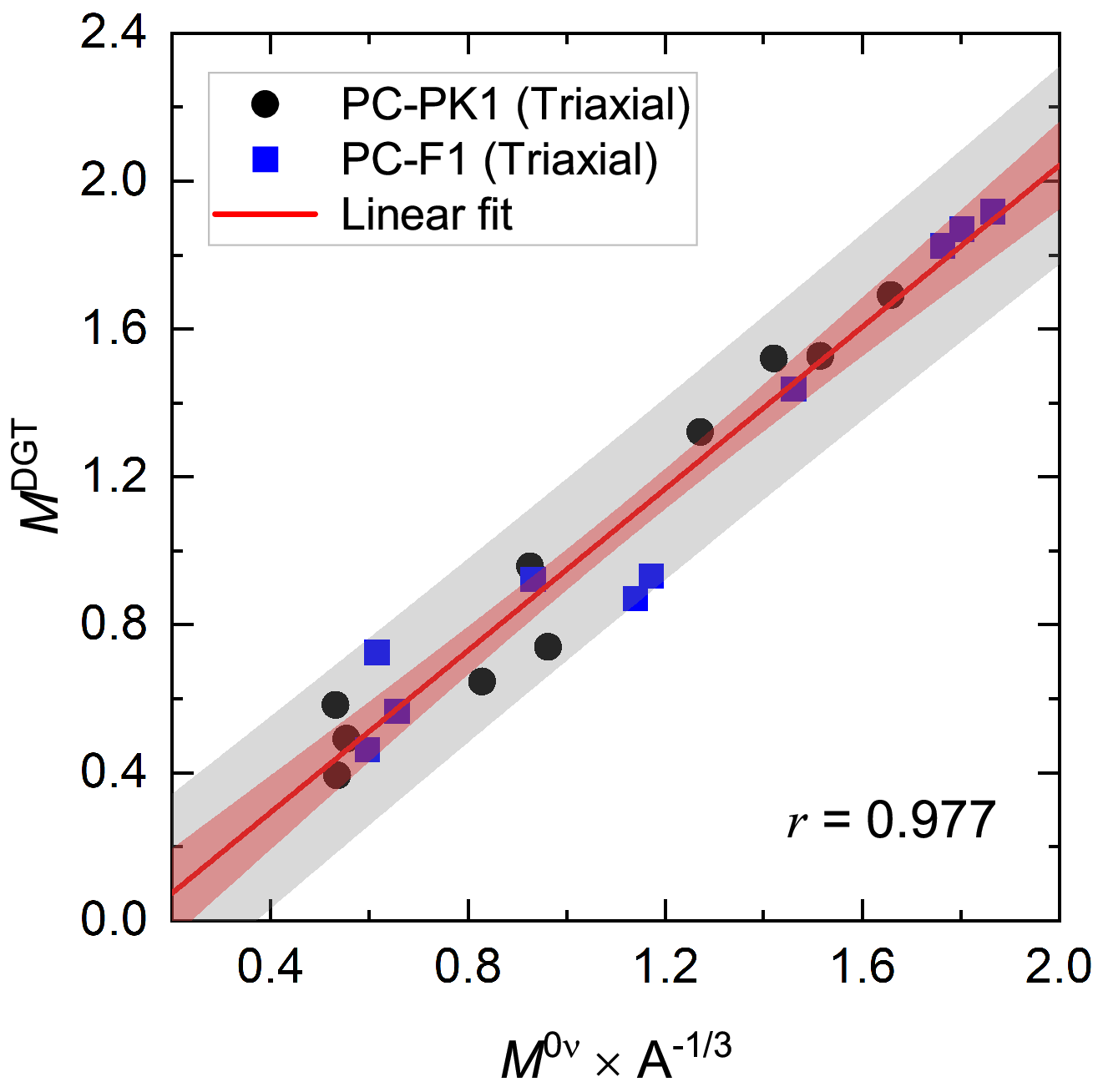}
  \caption{(Color online) Correlations between $M^{\mathrm{DGT}}$ and $M^{0\nu}$ calculated by the ReCD theory with PC-PK1 (dots) and PC-F1 (squares) density functionals.
  The $M^{0\nu}$ values are scaled by a factor $A^{-1/3}$.
  The inner (outer) colored regions depict the 95\% confidence (prediction) intervals of the linear regression, and the Pearson's correlation coefficient $r = 0.977$.}
  \label{Fig:Correlation-triaixal-axial}
\end{figure}
The calculated NMEs for the $0\nu\beta\beta$ decay $M^{0\nu}$ and the DGT transition $M^{\mathrm{DGT}}$ are depicted in Fig.~\ref{Fig:Correlation-triaixal-axial}.
A scaling factor $A^{-1/3}$ with $A$ the nuclear mass number is introduced to $M^{0\nu}$ to remove the mass-number dependence of the $0\nu\beta\beta$-decay NMEs~\cite{Yao2022Phys.Rev.C014315}.
A strong linear correlation between $M^{0\nu}$ and $M^{\mathrm{DGT}}$ is demonstrated, and the corresponding Pearson's correlation coefficient $r = 0.977$.

\begin{figure}[htbp]
  \centering
  \includegraphics[width=1.0\textwidth]{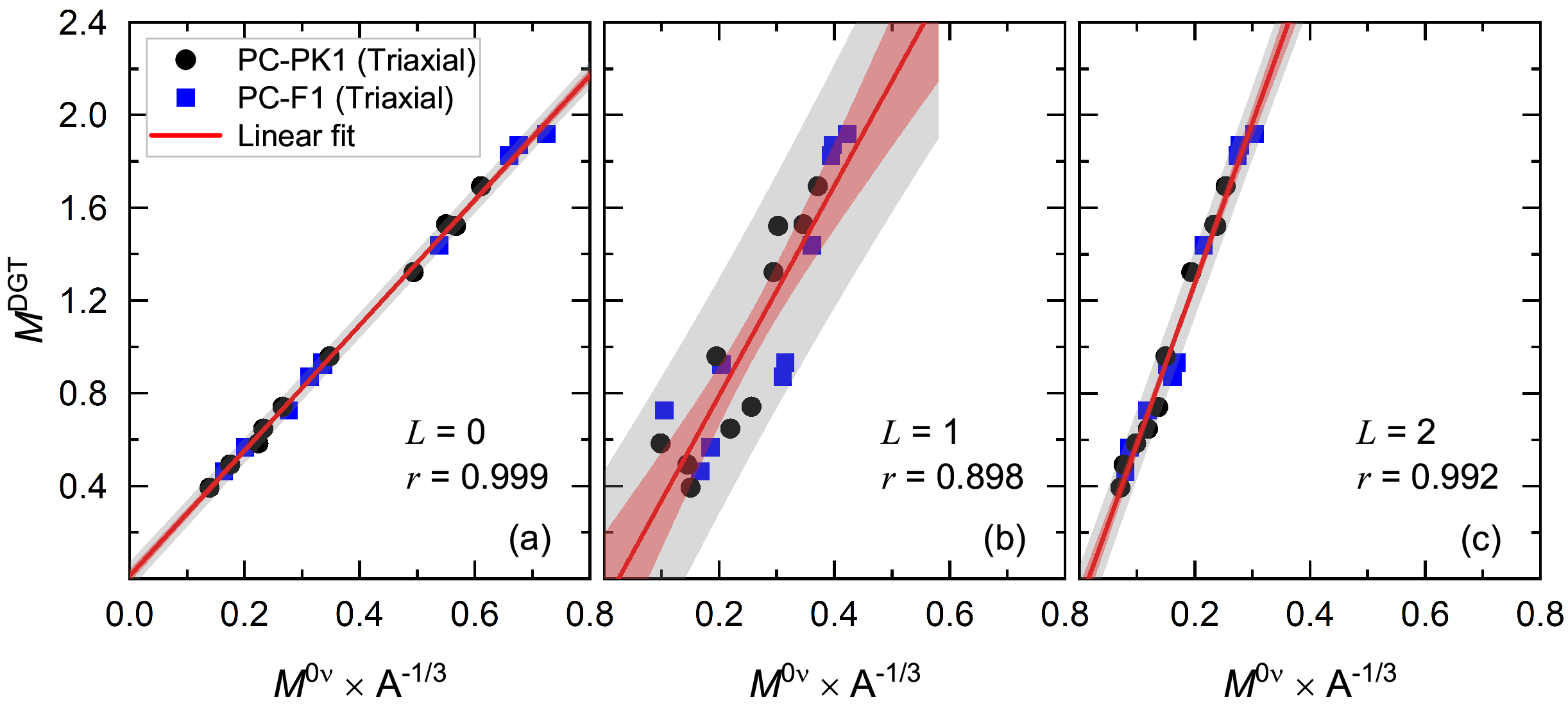}
  \caption{(Color online) Similar to that in Fig~\ref{Fig:Correlation-triaixal-axial}, but for correlations between $M^{\mathrm{DGT}}$ and $M^{0\nu}_{L}$ with $L = 0, 1, 2$.
  See text for more details.}
  \label{Fig:Correlation-L0-L1}
\end{figure}
To pin down the origin of the linear correlation, the $0\nu\beta\beta$-decay operator is expanded in terms of the spherical harmonics.
In this way, the $0\nu\beta\beta$-decay operator $\hat{O}^{0\nu}$ and the NME $M^{0\nu}$ can be expressed respectively as $\hat{O}^{0\nu} = \sum_L\hat{O}^{0\nu}_L$ and $M^{0\nu} = \sum_L M^{0\nu}_L$, with $L$ the rank of the spherical harmonics.
The obtained $M^{0\nu}_L$ with $L = 0, 1, 2$ and their correlations with $M^{\mathrm{DGT}}$ are shown in Fig.~\ref{Fig:Correlation-L0-L1}. 
It is clearly seen that the dominant leading-order term $M^{0\nu}_{L=0}$ strongly correlates with $M^{\mathrm{DGT}}$ and the Pearson's correlation coefficient $r = 0.999$.
The correlation between $M^{0\nu}_{L=1}$ and $M^{\mathrm{DGT}}$ is much weaker and the Pearson's correlation coefficient $r = 0.898$.
Compared to $M^{0\nu}_{L=0}$, $M^{0\nu}_L$ with $L = 2, 4, 6, \cdots$ are suppressed, and including their contributions would not worsen the correlation between the $0\nu\beta\beta$ decay and the DGT transition.
As an example, the correlation between $M^{0\nu}_{L=2}$ and $M^{\mathrm{DGT}}$ is shown in Fig.~\ref{Fig:Correlation-L0-L1} (c).
The consideration of higher-order terms with odd-$L$ values tend to contaminate the linear correlation.
However, the calculations show that $M^{0\nu}_L$ with odd $L$ values are smaller than $M^{0\nu}_L$ with even $L$ values.
These lead to the fact that the final $M^{0\nu}$, obtained by summing over $M^{0\nu}_L$ with all $L$ components, still correlates with $M^{\mathrm{DGT}}$.

The results presented in  Fig.~\ref{Fig:Correlation-L0-L1} suggest that dominant leading-order term $M^{0\nu}_{L=0}$ plays a key role for the presence of the strong correlation between $M^{0\nu}$ and $M^{\mathrm{DGT}}$.
In the following, the formula of decay operator $\hat{O}^{0\nu}_{L=0}$ corresponding to $M^{0\nu}_{L=0}$ will be derived in terms of the spherical harmonics.
Then, the similarity between  $\hat{O}^{0\nu}_{L=0}$ and DGT-transition operator $\hat{O}^{\mathrm{DGT}}$ will  be demonstrated to explain the strong correlation between $M^{0\nu}_{L=0}$  and $M^{\mathrm{DGT}}$.
Due to the fact that the AA coupling channel dominates the $0\nu\beta\beta$-decay process and its contribution exhausts more than 95\% of the total NME~\cite{Yao2015Phys.Rev.C024316}, we focus on the derivation of $\hat{O}^{0\nu}_{L=0}$ in the AA coupling channel. 

\begin{figure}[htbp]
  \centering
  \includegraphics[width=0.5\textwidth]{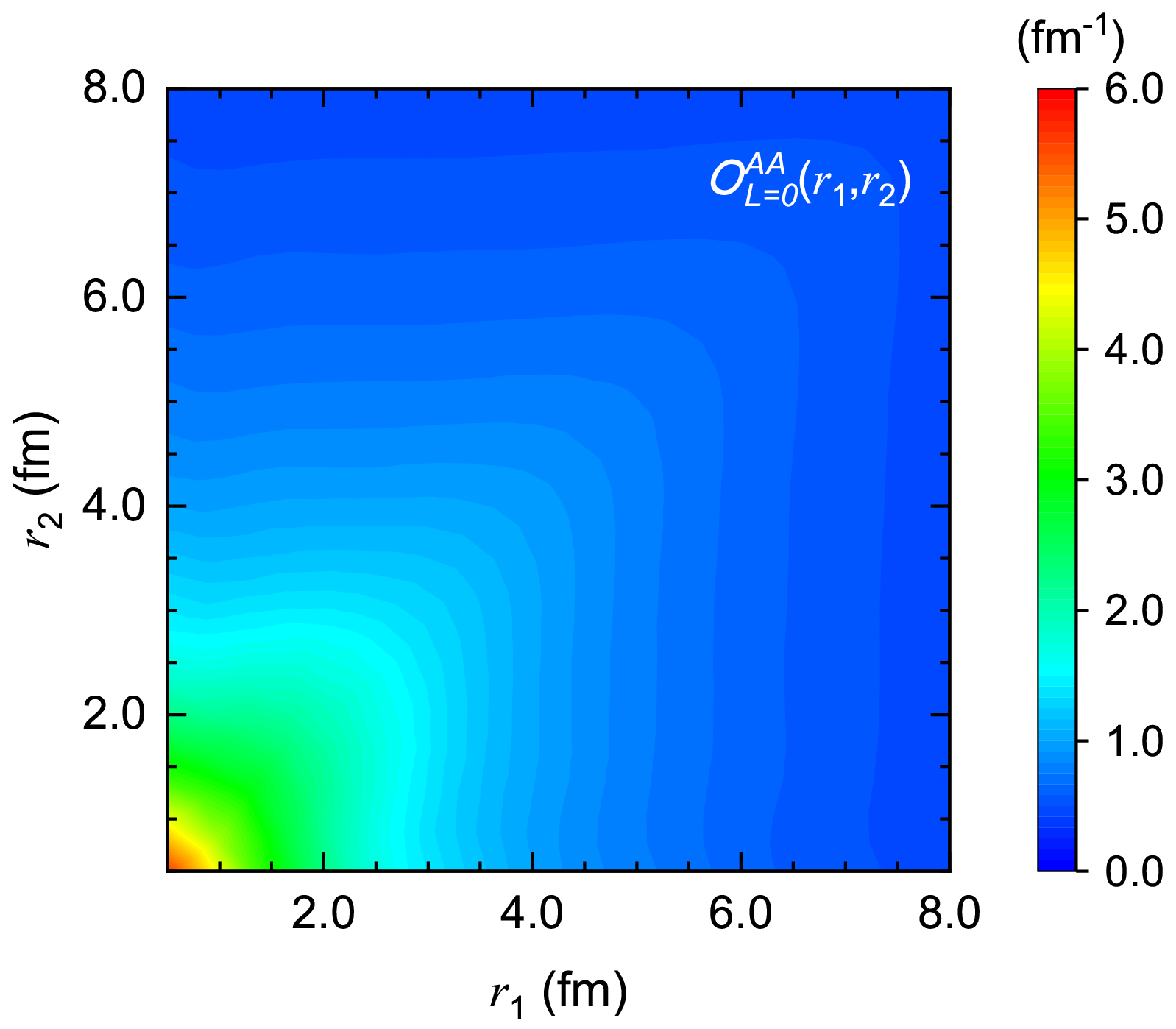}
  \caption{(Color online) Distribution of the neutrino potential $\mathcal{O}^{AA}_{L=0}(r_1,r_2)$ as functions of the nucleon coordinates $r_1$ and $r_2$.}
  \label{Fig:Neutrino-potential}
\end{figure}
The decay operator of the AA coupling channel in its second-quantized form reads
\begin{equation}\label{eq:AA-operator}
  \hat{O}^{0\nu}_{AA} = \sum_{1234} \langle 13|\mathcal{O}^{AA}(\bm{r}_1,\bm{r}_2)\bm{\sigma}_1\cdotp\bm{\sigma}_2|24\rangle\hat{d}^\dag_1\hat{d}^\dag_3\hat{c}_4\hat{c}_2,
\end{equation}
where $\hat{d}^\dag$ is the proton creation operator, $\hat{c}$ is the neutron annihilation operator, and indices 1, 2, 3, 4 characterize a set of spherical harmonic oscillator bases $|nljm\rangle$~\cite{Ring2004}.
The neutrino potential $\mathcal{O}^{AA}(\bm{r}_1,\bm{r}_2)$ in coordinate space is
\begin{equation}\label{eq:neutrino-pot}
  \mathcal{O}^{AA}(\bm{r}_1,\bm{r}_2) = \int \frac{d\bm{q}}{(2\pi)^3}H(\bm{q})e^{i\bm{q}\cdotp(\bm{r}_1-\bm{r}_2)},
\end{equation}
with $H(\bm{q})$ the neutrino potential in momentum space.
See more details about the decay operator in Refs.~\cite{Engel2017Rep.Prog.Phys.046301,Agostini2023Rev.Mod.Phys.025002}.
Using the multipole expansion for plane waves $e^{\pm i\bm{q}\cdotp\bm{r}}$ by the spherical harmonics, $\mathcal{O}^{AA}(\bm{r}_1,\bm{r}_2)$ can be reformulated as
\begin{equation}\label{eq:neutrino-pot1}
  \mathcal{O}^{AA}(\bm{r}_1,\bm{r}_2) = \frac{2}{\pi}\int q^2dq H(q) \sum_{LM} [j_L(qr_1)Y_{LM}(\hat{\bm{r}}_1)][j_L(qr_2)Y_{LM}^\ast(\hat{\bm{r}}_2)],
\end{equation}
where $j_L$ and $Y_{LM}$ denote respectively the spherical Bessel function and the spherical harmonics with rank $L$.
The neutrino potential with $L = 0$ in Eq.~\eqref{eq:neutrino-pot1} has the following form,
\begin{equation}\label{eq:neutrino-pot-L0}
  \mathcal{O}^{AA}_{L=0}(r_1,r_2) = \frac{1}{2\pi^2}\int q^2dqH(q)j_0(qr_1)j_0(qr_2),
\end{equation}
and its distribution as functions of nucleon coordinates $r_1$ and $r_2$ is depicted in Fig.~\ref{Fig:Neutrino-potential}.
It should be emphasized that there is no analytical expression for $\mathcal{O}^{AA}_{L=0}(r_1,r_2)$.
However, as a function of two variables, $\mathcal{O}^{AA}_{L=0}(r_1,r_2)$ can be decomposed as $\mathcal{O}^{AA}_{L=0}(r_1,r_2) = \sum_{ij}a_{ij}X_i(r_1)Y_j(r_2)$.
By integrating Eq.~\eqref{eq:neutrino-pot-L0} over the momentum $q$ numerically, it is found that $\mathcal{O}^{AA}_{L=0}(r_1,r_2)$ can be well approximated by
\begin{equation}\label{eq:neutrino-pot-L0-r}
  \mathcal{O}^{AA}_{L=0}(r_1,r_2) \approx \frac{1}{2}\left[X_1(r_1)Y_1(r_2) + X_1(r_2)Y_1(r_1)\right].
\end{equation}
Here, $X_1(r)$ and $Y_1(r)$ are smoothly decreasing functions that are larger than zero and do not have any node.
The single-particle matrix elements in terms of $|nljm\rangle$ then read approximately
\begin{equation}\label{eq:matrix-L0-nljm}
  \begin{split}
  \langle 13|\mathcal{O}^{AA}_{L=0}(r_1,r_2)\bm{\sigma}_1\cdotp\bm{\sigma}_2|24\rangle &\approx \langle n_1l_1|X_1(r_1)|n_2l_2\rangle\langle n_3l_3|Y_1(r_2)|n_4l_4\rangle\\
  &\times\langle j_1m_1|\bm{\sigma}_1|j_2m_2\rangle\langle j_3m_3|\bm{\sigma}_2|j_4m_4\rangle\delta_{n_1n_2}\delta_{n_3n_4}\delta_{l_1l_2}\delta_{l_3l_4}.
  \end{split}
\end{equation}
The contributions of $\langle n_1l_1|X_1(r_1)|n_2l_2\rangle$ and $\langle n_3l_3|Y_1(r_2)|n_4l_4\rangle$ with $n_1 \neq n_2$ or $n_3 \neq n_4$ are suppressed, as the radial nodes of the wavefunctions are different. 
Comparing Eq.~\eqref{eq:matrix-L0-nljm} to the single-particle matrix elements of the DGT-transition operator,
\begin{equation}\label{eq:matrix-GT-nljm}
  \langle 13|\mathcal{O}^{\mathrm{DGT}}|24\rangle = \frac{1}{\sqrt{3}}\langle j_1m_1|\bm{\sigma}_1|j_2m_2\rangle\langle j_3m_3|\bm{\sigma}_2|j_4m_4\rangle\delta_{n_1n_2}\delta_{n_3n_4}\delta_{l_1l_2}\delta_{l_3l_4},
\end{equation}
one can easily find that they share the same structure and, thus, the corresponding decay operators capture similar nuclear many-body correlations.
This explains the strong correlation between $M^{0\nu}_{L=0}$ and $M^{\mathrm{DGT}}$. 

\begin{figure}[htbp]
  \centering
  \includegraphics[width=0.5\textwidth]{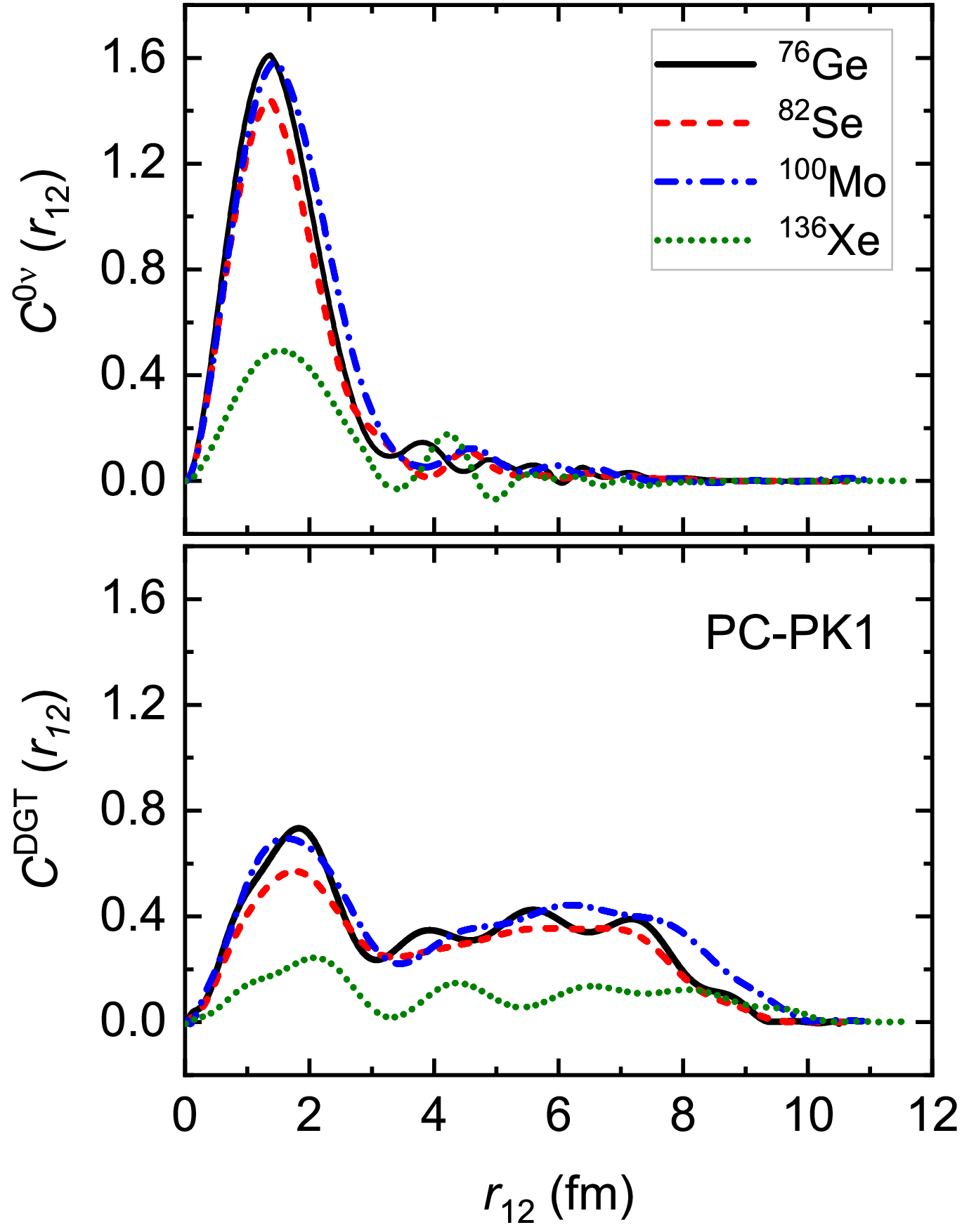}
  \caption{(Color online) The NME distributions for the $0\nu\beta\beta$ decay (upper) and the DGT transition (lower) as a function of relative distance $r_{12} = |\bm{r}_1 - \bm{r}_2|$ between the decaying nucleons.}
  \label{Fig:NME-distribution}
\end{figure}
We have noticed that the presence of the linear correlation was attributed to the dominant short-range character both in the $0\nu\beta\beta$ decay and the DGT transition in the previous study ~\cite{Shimizu2018Phys.Rev.Lett.142502}.
To examine such explanation, the NME distributions $C^{\alpha}(r_{12})$, defined by $\int dr_{12} C^{\alpha}(r_{12}) = M^\alpha$ with $\alpha$ specifying either $0\nu\beta\beta$ or DGT, are depicted in Fig.~\ref{Fig:NME-distribution}.
The results for $^{76}$Ge, $^{82}$Se, $^{100}$Mo, and $^{136}$Xe are shown as examples.
The $0\nu\beta\beta$-decay NMEs mainly distribute at the range of $r_{12} < 3$ fm, which corresponds to the so-called dominant short-range character of the $0\nu\beta\beta$ decay~\cite{Shimizu2018Phys.Rev.Lett.142502}.
However, the results of $C^{\mathrm{DGT}}(r_{12})$ indicate that the NMEs of the DGT transition are not short-range dominated.
Despite the distinct features of $C^{0\nu}$ and $C^{\mathrm{DGT}}$ as a function of $r_{12}$, the good linear correlation still appears.
Therefore, the explanation for the presence of the linear correlation given in Ref.~\cite{Shimizu2018Phys.Rev.Lett.142502} is not supported by the present study.

In summary, the NMEs of the $0\nu\beta\beta$ decay and the DGT transition in ten nuclei that are most relevant to the current and next-generation $0\nu\beta\beta$ decay experiments are investigated with the ReCD theory.
A strong linear correlation between the $0\nu\beta\beta$ decay and the DGT transition is demonstrated.
To understand the origin of the linear correlation, the $0\nu\beta\beta$-decay operator is expanded by using the spherical harmonics.
It is found that the dominant leading-order term of the $0\nu\beta\beta$-decay operator is very similar to the DGT-transition one, and this explains the presence of the linear correlation.
The present results provide a strong support to constrain the $0\nu\beta\beta$-decay NMEs from the double charge-exchange reactions.
To examine the conclusion in Ref.~\cite{Shimizu2018Phys.Rev.Lett.142502}, i.e., the linear correlation originates from the dominant short-range character, the NME distributions as a function of relative distance between two decaying nucleons are shown.
The so-called short-range dominant character is observed in the $0\nu\beta\beta$ decay, but not in the DGT transitions.

\begin{acknowledgments}
This work was partly supported by the National Natural Science Foundation of China (Grants No. 12141501, No. 12105004, No. 12070131001, No. 11875075, No. 11935003, and No. 11975031), the China Postdoctoral Science Foundation under Grant No. 2020M680183, the State Key Laboratory of Nuclear Physics and Technology, Peking University, and the High-performance Computing Platform of Peking University.
\end{acknowledgments}


\begin{thebibliography}{37}%
\makeatletter
\providecommand \@ifxundefined [1]{%
 \@ifx{#1\undefined}
}%
\providecommand \@ifnum [1]{%
 \ifnum #1\expandafter \@firstoftwo
 \else \expandafter \@secondoftwo
 \fi
}%
\providecommand \@ifx [1]{%
 \ifx #1\expandafter \@firstoftwo
 \else \expandafter \@secondoftwo
 \fi
}%
\providecommand \natexlab [1]{#1}%
\providecommand \enquote  [1]{``#1''}%
\providecommand \bibnamefont  [1]{#1}%
\providecommand \bibfnamefont [1]{#1}%
\providecommand \citenamefont [1]{#1}%
\providecommand \href@noop [0]{\@secondoftwo}%
\providecommand \href [0]{\begingroup \@sanitize@url \@href}%
\providecommand \@href[1]{\@@startlink{#1}\@@href}%
\providecommand \@@href[1]{\endgroup#1\@@endlink}%
\providecommand \@sanitize@url [0]{\catcode `\\12\catcode `\$12\catcode
  `\&12\catcode `\#12\catcode `\^12\catcode `\_12\catcode `\%12\relax}%
\providecommand \@@startlink[1]{}%
\providecommand \@@endlink[0]{}%
\providecommand \url  [0]{\begingroup\@sanitize@url \@url }%
\providecommand \@url [1]{\endgroup\@href {#1}{\urlprefix }}%
\providecommand \urlprefix  [0]{URL }%
\providecommand \Eprint [0]{\href }%
\providecommand \doibase [0]{https://doi.org/}%
\providecommand \selectlanguage [0]{\@gobble}%
\providecommand \bibinfo  [0]{\@secondoftwo}%
\providecommand \bibfield  [0]{\@secondoftwo}%
\providecommand \translation [1]{[#1]}%
\providecommand \BibitemOpen [0]{}%
\providecommand \bibitemStop [0]{}%
\providecommand \bibitemNoStop [0]{.\EOS\space}%
\providecommand \EOS [0]{\spacefactor3000\relax}%
\providecommand \BibitemShut  [1]{\csname bibitem#1\endcsname}%
\let\auto@bib@innerbib\@empty
\bibitem [{\citenamefont {Aalseth}\ \emph {et~al.}(2018)\citenamefont
  {Aalseth}, \citenamefont {Abgrall}, \citenamefont {Aguayo}, \citenamefont
  {Alvis}, \citenamefont {Amman}, \citenamefont {Arnquist}, \citenamefont
  {Avignone}, \citenamefont {Back}, \citenamefont {Barabash}, \citenamefont
  {Barbeau} \emph {et~al.}}]{Aalseth2018Phys.Rev.Lett.132502}%
  \BibitemOpen
  \bibfield  {author} {\bibinfo {author} {\bibfnamefont {C.~E.}\ \bibnamefont
  {Aalseth}}, \bibinfo {author} {\bibfnamefont {N.}~\bibnamefont {Abgrall}},
  \bibinfo {author} {\bibfnamefont {E.}~\bibnamefont {Aguayo}}, \bibinfo
  {author} {\bibfnamefont {S.~I.}\ \bibnamefont {Alvis}}, \bibinfo {author}
  {\bibfnamefont {M.}~\bibnamefont {Amman}}, \bibinfo {author} {\bibfnamefont
  {I.~J.}\ \bibnamefont {Arnquist}}, \bibinfo {author} {\bibfnamefont {F.~T.}\
  \bibnamefont {Avignone}}, \bibinfo {author} {\bibfnamefont {H.~O.}\
  \bibnamefont {Back}}, \bibinfo {author} {\bibfnamefont {A.~S.}\ \bibnamefont
  {Barabash}}, \bibinfo {author} {\bibfnamefont {P.~S.}\ \bibnamefont
  {Barbeau}}, \emph {et~al.} (\bibinfo {collaboration} {Majorana
  Collaboration}),\ }\bibfield  {title} {\bibinfo {title} {Search for
  neutrinoless double-$\ensuremath{\beta}$ decay in $^{76}\mathrm{Ge}$ with the
  majorana demonstrator},\ }\href
  {https://doi.org/10.1103/PhysRevLett.120.132502} {\bibfield  {journal}
  {\bibinfo  {journal} {Phys. Rev. Lett.}\ }\textbf {\bibinfo {volume} {120}},\
  \bibinfo {pages} {132502} (\bibinfo {year} {2018})}\BibitemShut {NoStop}%
\bibitem [{\citenamefont {Adams}\ \emph {et~al.}(2020)\citenamefont {Adams},
  \citenamefont {Alduino}, \citenamefont {Alfonso}, \citenamefont {Avignone},
  \citenamefont {Azzolini}, \citenamefont {Bari}, \citenamefont {Bellini},
  \citenamefont {Benato}, \citenamefont {Biassoni} \emph
  {et~al.}}]{Adams2020Phys.Rev.Lett.122501}%
  \BibitemOpen
  \bibfield  {author} {\bibinfo {author} {\bibfnamefont {D.~Q.}\ \bibnamefont
  {Adams}}, \bibinfo {author} {\bibfnamefont {C.}~\bibnamefont {Alduino}},
  \bibinfo {author} {\bibfnamefont {K.}~\bibnamefont {Alfonso}}, \bibinfo
  {author} {\bibfnamefont {F.~T.}\ \bibnamefont {Avignone}}, \bibinfo {author}
  {\bibfnamefont {O.}~\bibnamefont {Azzolini}}, \bibinfo {author}
  {\bibfnamefont {G.}~\bibnamefont {Bari}}, \bibinfo {author} {\bibfnamefont
  {F.}~\bibnamefont {Bellini}}, \bibinfo {author} {\bibfnamefont
  {G.}~\bibnamefont {Benato}}, \bibinfo {author} {\bibfnamefont
  {M.}~\bibnamefont {Biassoni}}, \emph {et~al.} (\bibinfo {collaboration}
  {CUORE Collaboration}),\ }\bibfield  {title} {\bibinfo {title} {Improved
  limit on neutrinoless double-beta decay in $^{130} \mathrm{Te}$ with cuore},\
  }\href {https://doi.org/10.1103/PhysRevLett.124.122501} {\bibfield  {journal}
  {\bibinfo  {journal} {Phys. Rev. Lett.}\ }\textbf {\bibinfo {volume} {124}},\
  \bibinfo {pages} {122501} (\bibinfo {year} {2020})}\BibitemShut {NoStop}%
\bibitem [{\citenamefont {Agostini}\ \emph {et~al.}(2020)\citenamefont
  {Agostini}, \citenamefont {Araujo}, \citenamefont {Bakalyarov}, \citenamefont
  {Balata}, \citenamefont {Barabanov}, \citenamefont {Baudis}, \citenamefont
  {Bauer}, \citenamefont {Bellotti}, \citenamefont {Belogurov}, \citenamefont
  {Bettini} \emph {et~al.}}]{Agostini2020Phys.Rev.Lett.252502}%
  \BibitemOpen
  \bibfield  {author} {\bibinfo {author} {\bibfnamefont {M.}~\bibnamefont
  {Agostini}}, \bibinfo {author} {\bibfnamefont {G.~R.}\ \bibnamefont
  {Araujo}}, \bibinfo {author} {\bibfnamefont {A.~M.}\ \bibnamefont
  {Bakalyarov}}, \bibinfo {author} {\bibfnamefont {M.}~\bibnamefont {Balata}},
  \bibinfo {author} {\bibfnamefont {I.}~\bibnamefont {Barabanov}}, \bibinfo
  {author} {\bibfnamefont {L.}~\bibnamefont {Baudis}}, \bibinfo {author}
  {\bibfnamefont {C.}~\bibnamefont {Bauer}}, \bibinfo {author} {\bibfnamefont
  {E.}~\bibnamefont {Bellotti}}, \bibinfo {author} {\bibfnamefont
  {S.}~\bibnamefont {Belogurov}}, \bibinfo {author} {\bibfnamefont
  {A.}~\bibnamefont {Bettini}}, \emph {et~al.} (\bibinfo {collaboration} {GERDA
  Collaboration}),\ }\bibfield  {title} {\bibinfo {title} {Final results of
  gerda on the search for neutrinoless double-$\ensuremath{\beta}$ decay},\
  }\href {https://doi.org/10.1103/PhysRevLett.125.252502} {\bibfield  {journal}
  {\bibinfo  {journal} {Phys. Rev. Lett.}\ }\textbf {\bibinfo {volume} {125}},\
  \bibinfo {pages} {252502} (\bibinfo {year} {2020})}\BibitemShut {NoStop}%
\bibitem [{\citenamefont {Albert}\ \emph {et~al.}(2018)\citenamefont {Albert},
  \citenamefont {Anton}, \citenamefont {Badhrees}, \citenamefont {Barbeau},
  \citenamefont {Bayerlein}, \citenamefont {Beck}, \citenamefont {Belov},
  \citenamefont {Breidenbach}, \citenamefont {Brunner}, \citenamefont {Cao}
  \emph {et~al.}}]{Albert2018Phys.Rev.Lett.072701}%
  \BibitemOpen
  \bibfield  {author} {\bibinfo {author} {\bibfnamefont {J.~B.}\ \bibnamefont
  {Albert}}, \bibinfo {author} {\bibfnamefont {G.}~\bibnamefont {Anton}},
  \bibinfo {author} {\bibfnamefont {I.}~\bibnamefont {Badhrees}}, \bibinfo
  {author} {\bibfnamefont {P.~S.}\ \bibnamefont {Barbeau}}, \bibinfo {author}
  {\bibfnamefont {R.}~\bibnamefont {Bayerlein}}, \bibinfo {author}
  {\bibfnamefont {D.}~\bibnamefont {Beck}}, \bibinfo {author} {\bibfnamefont
  {V.}~\bibnamefont {Belov}}, \bibinfo {author} {\bibfnamefont
  {M.}~\bibnamefont {Breidenbach}}, \bibinfo {author} {\bibfnamefont
  {T.}~\bibnamefont {Brunner}}, \bibinfo {author} {\bibfnamefont {G.~F.}\
  \bibnamefont {Cao}}, \emph {et~al.} (\bibinfo {collaboration} {EXO-200
  Collaboration}),\ }\bibfield  {title} {\bibinfo {title} {Search for
  neutrinoless double-beta decay with the upgraded exo-200 detector},\ }\href
  {https://doi.org/10.1103/PhysRevLett.120.072701} {\bibfield  {journal}
  {\bibinfo  {journal} {Phys. Rev. Lett.}\ }\textbf {\bibinfo {volume} {120}},\
  \bibinfo {pages} {072701} (\bibinfo {year} {2018})}\BibitemShut {NoStop}%
\bibitem [{\citenamefont {Armengaud}\ \emph {et~al.}(2021)\citenamefont
  {Armengaud}, \citenamefont {Augier}, \citenamefont {Barabash}, \citenamefont
  {Bellini}, \citenamefont {Benato}, \citenamefont {Beno\^{\i}t}, \citenamefont
  {Beretta}, \citenamefont {Berg\'e}, \citenamefont {Billard}, \citenamefont
  {Borovlev} \emph {et~al.}}]{Armengaud2021Phys.Rev.Lett.181802}%
  \BibitemOpen
  \bibfield  {author} {\bibinfo {author} {\bibfnamefont {E.}~\bibnamefont
  {Armengaud}}, \bibinfo {author} {\bibfnamefont {C.}~\bibnamefont {Augier}},
  \bibinfo {author} {\bibfnamefont {A.~S.}\ \bibnamefont {Barabash}}, \bibinfo
  {author} {\bibfnamefont {F.}~\bibnamefont {Bellini}}, \bibinfo {author}
  {\bibfnamefont {G.}~\bibnamefont {Benato}}, \bibinfo {author} {\bibfnamefont
  {A.}~\bibnamefont {Beno\^{\i}t}}, \bibinfo {author} {\bibfnamefont
  {M.}~\bibnamefont {Beretta}}, \bibinfo {author} {\bibfnamefont
  {L.}~\bibnamefont {Berg\'e}}, \bibinfo {author} {\bibfnamefont
  {J.}~\bibnamefont {Billard}}, \bibinfo {author} {\bibfnamefont {Y.~A.}\
  \bibnamefont {Borovlev}}, \emph {et~al.} (\bibinfo {collaboration} {CUPID-Mo
  Collaboration}),\ }\bibfield  {title} {\bibinfo {title} {New limit for
  neutrinoless double-beta decay of $^{100}\mathrm{Mo}$ from the cupid-mo
  experiment},\ }\href {https://doi.org/10.1103/PhysRevLett.126.181802}
  {\bibfield  {journal} {\bibinfo  {journal} {Phys. Rev. Lett.}\ }\textbf
  {\bibinfo {volume} {126}},\ \bibinfo {pages} {181802} (\bibinfo {year}
  {2021})}\BibitemShut {NoStop}%
\bibitem [{\citenamefont {Arnold}\ \emph {et~al.}(2017)\citenamefont {Arnold},
  \citenamefont {Augier}, \citenamefont {Barabash}, \citenamefont
  {Basharina-Freshville}, \citenamefont {Blondel}, \citenamefont {Blot},
  \citenamefont {Bongrand}, \citenamefont {Boursette}, \citenamefont
  {Brudanin}, \citenamefont {Busto} \emph
  {et~al.}}]{Arnold2017Phys.Rev.Lett.041801}%
  \BibitemOpen
  \bibfield  {author} {\bibinfo {author} {\bibfnamefont {R.}~\bibnamefont
  {Arnold}}, \bibinfo {author} {\bibfnamefont {C.}~\bibnamefont {Augier}},
  \bibinfo {author} {\bibfnamefont {A.~S.}\ \bibnamefont {Barabash}}, \bibinfo
  {author} {\bibfnamefont {A.}~\bibnamefont {Basharina-Freshville}}, \bibinfo
  {author} {\bibfnamefont {S.}~\bibnamefont {Blondel}}, \bibinfo {author}
  {\bibfnamefont {S.}~\bibnamefont {Blot}}, \bibinfo {author} {\bibfnamefont
  {M.}~\bibnamefont {Bongrand}}, \bibinfo {author} {\bibfnamefont
  {D.}~\bibnamefont {Boursette}}, \bibinfo {author} {\bibfnamefont
  {V.}~\bibnamefont {Brudanin}}, \bibinfo {author} {\bibfnamefont
  {J.}~\bibnamefont {Busto}}, \emph {et~al.} (\bibinfo {collaboration} {NEMO-3
  Collaboration}),\ }\bibfield  {title} {\bibinfo {title} {Search for
  neutrinoless quadruple-$\ensuremath{\beta}$ decay of $^{150}\mathrm{Nd}$ with
  the nemo-3 detector},\ }\href
  {https://doi.org/10.1103/PhysRevLett.119.041801} {\bibfield  {journal}
  {\bibinfo  {journal} {Phys. Rev. Lett.}\ }\textbf {\bibinfo {volume} {119}},\
  \bibinfo {pages} {041801} (\bibinfo {year} {2017})}\BibitemShut {NoStop}%
\bibitem [{\citenamefont {Gando}\ \emph {et~al.}(2016)\citenamefont {Gando},
  \citenamefont {Gando}, \citenamefont {Hachiya}, \citenamefont {Hayashi},
  \citenamefont {Hayashida}, \citenamefont {Ikeda}, \citenamefont {Inoue},
  \citenamefont {Ishidoshiro}, \citenamefont {Karino}, \citenamefont {Koga}
  \emph {et~al.}}]{Gando2016Phys.Rev.Lett.082503}%
  \BibitemOpen
  \bibfield  {author} {\bibinfo {author} {\bibfnamefont {A.}~\bibnamefont
  {Gando}}, \bibinfo {author} {\bibfnamefont {Y.}~\bibnamefont {Gando}},
  \bibinfo {author} {\bibfnamefont {T.}~\bibnamefont {Hachiya}}, \bibinfo
  {author} {\bibfnamefont {A.}~\bibnamefont {Hayashi}}, \bibinfo {author}
  {\bibfnamefont {S.}~\bibnamefont {Hayashida}}, \bibinfo {author}
  {\bibfnamefont {H.}~\bibnamefont {Ikeda}}, \bibinfo {author} {\bibfnamefont
  {K.}~\bibnamefont {Inoue}}, \bibinfo {author} {\bibfnamefont
  {K.}~\bibnamefont {Ishidoshiro}}, \bibinfo {author} {\bibfnamefont
  {Y.}~\bibnamefont {Karino}}, \bibinfo {author} {\bibfnamefont
  {M.}~\bibnamefont {Koga}}, \emph {et~al.} (\bibinfo {collaboration}
  {KamLAND-Zen Collaboration}),\ }\bibfield  {title} {\bibinfo {title} {Search
  for majorana neutrinos near the inverted mass hierarchy region with
  kamland-zen},\ }\href {https://doi.org/10.1103/PhysRevLett.117.082503}
  {\bibfield  {journal} {\bibinfo  {journal} {Phys. Rev. Lett.}\ }\textbf
  {\bibinfo {volume} {117}},\ \bibinfo {pages} {082503} (\bibinfo {year}
  {2016})}\BibitemShut {NoStop}%
\bibitem [{\citenamefont {Dai}\ \emph {et~al.}(2022)\citenamefont {Dai},
  \citenamefont {Ma}, \citenamefont {Yue}, \citenamefont {She}, \citenamefont
  {Kang}, \citenamefont {Li}, \citenamefont {Agartioglu}, \citenamefont {An},
  \citenamefont {Chang}, \citenamefont {Chen} \emph
  {et~al.}}]{Dai2022Phys.Rev.D032012}%
  \BibitemOpen
  \bibfield  {author} {\bibinfo {author} {\bibfnamefont {W.~H.}\ \bibnamefont
  {Dai}}, \bibinfo {author} {\bibfnamefont {H.}~\bibnamefont {Ma}}, \bibinfo
  {author} {\bibfnamefont {Q.}~\bibnamefont {Yue}}, \bibinfo {author}
  {\bibfnamefont {Z.}~\bibnamefont {She}}, \bibinfo {author} {\bibfnamefont
  {K.~J.}\ \bibnamefont {Kang}}, \bibinfo {author} {\bibfnamefont {Y.~J.}\
  \bibnamefont {Li}}, \bibinfo {author} {\bibfnamefont {M.}~\bibnamefont
  {Agartioglu}}, \bibinfo {author} {\bibfnamefont {H.~P.}\ \bibnamefont {An}},
  \bibinfo {author} {\bibfnamefont {J.~P.}\ \bibnamefont {Chang}}, \bibinfo
  {author} {\bibfnamefont {Y.~H.}\ \bibnamefont {Chen}}, \emph {et~al.}
  (\bibinfo {collaboration} {CDEX Collaboration}),\ }\bibfield  {title}
  {\bibinfo {title} {Search for neutrinoless double-beta decay of
  $^{76}\mathrm{Ge}$ with a natural broad energy germanium detector},\ }\href
  {https://doi.org/10.1103/PhysRevD.106.032012} {\bibfield  {journal} {\bibinfo
   {journal} {Phys. Rev. D}\ }\textbf {\bibinfo {volume} {106}},\ \bibinfo
  {pages} {032012} (\bibinfo {year} {2022})}\BibitemShut {NoStop}%
\bibitem [{\citenamefont {Engel}\ and\ \citenamefont
  {Menéndez}(2017)}]{Engel2017Rep.Prog.Phys.046301}%
  \BibitemOpen
  \bibfield  {author} {\bibinfo {author} {\bibfnamefont {J.}~\bibnamefont
  {Engel}}\ and\ \bibinfo {author} {\bibfnamefont {J.}~\bibnamefont
  {Menéndez}},\ }\bibfield  {title} {\bibinfo {title} {Status and future of
  nuclear matrix elements for neutrinoless double-beta decay: a review},\
  }\href {https://doi.org/10.1088/1361-6633/aa5bc5} {\bibfield  {journal}
  {\bibinfo  {journal} {Rep. Prog. Phys.}\ }\textbf {\bibinfo {volume} {80}},\
  \bibinfo {pages} {046301} (\bibinfo {year} {2017})}\BibitemShut {NoStop}%
\bibitem [{\citenamefont {Yao}\ \emph {et~al.}(2022{\natexlab{a}})\citenamefont
  {Yao}, \citenamefont {Meng}, \citenamefont {Niu},\ and\ \citenamefont
  {Ring}}]{Yao2022ProgressinParticleandNuclearPhysics103965}%
  \BibitemOpen
  \bibfield  {author} {\bibinfo {author} {\bibfnamefont {J.}~\bibnamefont
  {Yao}}, \bibinfo {author} {\bibfnamefont {J.}~\bibnamefont {Meng}}, \bibinfo
  {author} {\bibfnamefont {Y.}~\bibnamefont {Niu}},\ and\ \bibinfo {author}
  {\bibfnamefont {P.}~\bibnamefont {Ring}},\ }\bibfield  {title} {\bibinfo
  {title} {Beyond-mean-field approaches for nuclear neutrinoless double beta
  decay in the standard mechanism},\ }\href
  {https://doi.org/https://doi.org/10.1016/j.ppnp.2022.103965} {\bibfield
  {journal} {\bibinfo  {journal} {Prog. Part. Nucl. Phys.}\ }\textbf {\bibinfo
  {volume} {126}},\ \bibinfo {pages} {103965} (\bibinfo {year}
  {2022}{\natexlab{a}})}\BibitemShut {NoStop}%
\bibitem [{\citenamefont {Agostini}\ \emph {et~al.}(2023)\citenamefont
  {Agostini}, \citenamefont {Benato}, \citenamefont {Detwiler}, \citenamefont
  {Men\'endez},\ and\ \citenamefont
  {Vissani}}]{Agostini2023Rev.Mod.Phys.025002}%
  \BibitemOpen
  \bibfield  {author} {\bibinfo {author} {\bibfnamefont {M.}~\bibnamefont
  {Agostini}}, \bibinfo {author} {\bibfnamefont {G.}~\bibnamefont {Benato}},
  \bibinfo {author} {\bibfnamefont {J.~A.}\ \bibnamefont {Detwiler}}, \bibinfo
  {author} {\bibfnamefont {J.}~\bibnamefont {Men\'endez}},\ and\ \bibinfo
  {author} {\bibfnamefont {F.}~\bibnamefont {Vissani}},\ }\bibfield  {title}
  {\bibinfo {title} {Toward the discovery of matter creation with neutrinoless
  $\ensuremath{\beta}\ensuremath{\beta}$ decay},\ }\href
  {https://doi.org/10.1103/RevModPhys.95.025002} {\bibfield  {journal}
  {\bibinfo  {journal} {Rev. Mod. Phys.}\ }\textbf {\bibinfo {volume} {95}},\
  \bibinfo {pages} {025002} (\bibinfo {year} {2023})}\BibitemShut {NoStop}%
\bibitem [{\citenamefont {Rodr{\'\i}guez}\ and\ \citenamefont
  {Mart{\'\i}nez-Pinedo}(2013)}]{Rodriguez2013PhysicsLettersB174178}%
  \BibitemOpen
  \bibfield  {author} {\bibinfo {author} {\bibfnamefont {T.~R.}\ \bibnamefont
  {Rodr{\'\i}guez}}\ and\ \bibinfo {author} {\bibfnamefont {G.}~\bibnamefont
  {Mart{\'\i}nez-Pinedo}},\ }\bibfield  {title} {\bibinfo {title} {Neutrinoless
  $\beta$$\beta$ decay nuclear matrix elements in an isotopic chain},\
  }\href@noop {} {\bibfield  {journal} {\bibinfo  {journal} {Phys. Lett. B}\
  }\textbf {\bibinfo {volume} {719}},\ \bibinfo {pages} {174} (\bibinfo {year}
  {2013})}\BibitemShut {NoStop}%
\bibitem [{\citenamefont {Cappuzzello}\ \emph {et~al.}(2015)\citenamefont
  {Cappuzzello}, \citenamefont {Cavallaro}, \citenamefont {Agodi},
  \citenamefont {Bond{\`\i}}, \citenamefont {Carbone}, \citenamefont
  {Cunsolo},\ and\ \citenamefont
  {Foti}}]{Cappuzzello2015TheEuropeanPhysicalJournalA145}%
  \BibitemOpen
  \bibfield  {author} {\bibinfo {author} {\bibfnamefont {F.}~\bibnamefont
  {Cappuzzello}}, \bibinfo {author} {\bibfnamefont {M.}~\bibnamefont
  {Cavallaro}}, \bibinfo {author} {\bibfnamefont {C.}~\bibnamefont {Agodi}},
  \bibinfo {author} {\bibfnamefont {M.}~\bibnamefont {Bond{\`\i}}}, \bibinfo
  {author} {\bibfnamefont {D.}~\bibnamefont {Carbone}}, \bibinfo {author}
  {\bibfnamefont {A.}~\bibnamefont {Cunsolo}},\ and\ \bibinfo {author}
  {\bibfnamefont {A.}~\bibnamefont {Foti}},\ }\bibfield  {title} {\bibinfo
  {title} {Heavy-ion double charge exchange reactions: A tool toward nuclear
  matrix elements},\ }\href@noop {} {\bibfield  {journal} {\bibinfo  {journal}
  {Eur. Phys. Jour. A}\ }\textbf {\bibinfo {volume} {51}},\ \bibinfo {pages}
  {145} (\bibinfo {year} {2015})}\BibitemShut {NoStop}%
\bibitem [{\citenamefont {Takahisa}\ \emph {et~al.}(2017)\citenamefont
  {Takahisa}, \citenamefont {Ejiri}, \citenamefont {Akimune}, \citenamefont
  {Fujita}, \citenamefont {Matumiya}, \citenamefont {Ohta}, \citenamefont
  {Shima}, \citenamefont {Tanaka},\ and\ \citenamefont
  {Yosoi}}]{takahisa2017double}%
  \BibitemOpen
  \bibfield  {author} {\bibinfo {author} {\bibfnamefont {K.}~\bibnamefont
  {Takahisa}}, \bibinfo {author} {\bibfnamefont {H.}~\bibnamefont {Ejiri}},
  \bibinfo {author} {\bibfnamefont {H.}~\bibnamefont {Akimune}}, \bibinfo
  {author} {\bibfnamefont {H.}~\bibnamefont {Fujita}}, \bibinfo {author}
  {\bibfnamefont {R.}~\bibnamefont {Matumiya}}, \bibinfo {author}
  {\bibfnamefont {T.}~\bibnamefont {Ohta}}, \bibinfo {author} {\bibfnamefont
  {T.}~\bibnamefont {Shima}}, \bibinfo {author} {\bibfnamefont
  {M.}~\bibnamefont {Tanaka}},\ and\ \bibinfo {author} {\bibfnamefont
  {M.}~\bibnamefont {Yosoi}},\ }\href@noop {} {\bibfield  {journal} {\bibinfo
  {journal} {arXiv:1703.08264}\ } (\bibinfo {year} {2017})}\BibitemShut
  {NoStop}%
\bibitem [{\citenamefont {Santopinto}\ \emph {et~al.}(2018)\citenamefont
  {Santopinto}, \citenamefont {Garc\'{\i}a-Tecocoatzi}, \citenamefont {Maga\~na
  Vsevolodovna},\ and\ \citenamefont
  {Ferretti}}]{Santopinto2018Phys.Rev.C061601}%
  \BibitemOpen
  \bibfield  {author} {\bibinfo {author} {\bibfnamefont {E.}~\bibnamefont
  {Santopinto}}, \bibinfo {author} {\bibfnamefont {H.}~\bibnamefont
  {Garc\'{\i}a-Tecocoatzi}}, \bibinfo {author} {\bibfnamefont {R.~I.}\
  \bibnamefont {Maga\~na Vsevolodovna}},\ and\ \bibinfo {author} {\bibfnamefont
  {J.}~\bibnamefont {Ferretti}} (\bibinfo {collaboration} {NUMEN
  Collaboration}),\ }\bibfield  {title} {\bibinfo {title} {Heavy-ion
  double-charge-exchange and its relation to neutrinoless
  double-$\ensuremath{\beta}$ decay},\ }\href
  {https://doi.org/10.1103/PhysRevC.98.061601} {\bibfield  {journal} {\bibinfo
  {journal} {Phys. Rev. C}\ }\textbf {\bibinfo {volume} {98}},\ \bibinfo
  {pages} {061601} (\bibinfo {year} {2018})}\BibitemShut {NoStop}%
\bibitem [{\citenamefont {Shimizu}\ \emph {et~al.}(2018)\citenamefont
  {Shimizu}, \citenamefont {Men\'endez},\ and\ \citenamefont
  {Yako}}]{Shimizu2018Phys.Rev.Lett.142502}%
  \BibitemOpen
  \bibfield  {author} {\bibinfo {author} {\bibfnamefont {N.}~\bibnamefont
  {Shimizu}}, \bibinfo {author} {\bibfnamefont {J.}~\bibnamefont
  {Men\'endez}},\ and\ \bibinfo {author} {\bibfnamefont {K.}~\bibnamefont
  {Yako}},\ }\bibfield  {title} {\bibinfo {title} {Double gamow-teller
  transitions and its relation to neutrinoless
  $\ensuremath{\beta}\ensuremath{\beta}$ decay},\ }\href
  {https://doi.org/10.1103/PhysRevLett.120.142502} {\bibfield  {journal}
  {\bibinfo  {journal} {Phys. Rev. Lett.}\ }\textbf {\bibinfo {volume} {120}},\
  \bibinfo {pages} {142502} (\bibinfo {year} {2018})}\BibitemShut {NoStop}%
\bibitem [{\citenamefont {\ifmmode~\check{S}\else \v{S}\fi{}imkovic}\ \emph
  {et~al.}(2018)\citenamefont {\ifmmode~\check{S}\else \v{S}\fi{}imkovic},
  \citenamefont {Smetana},\ and\ \citenamefont
  {Vogel}}]{ifmmodeSelseSfiimkovic2018Phys.Rev.C064325}%
  \BibitemOpen
  \bibfield  {author} {\bibinfo {author} {\bibfnamefont {F.}~\bibnamefont
  {\ifmmode~\check{S}\else \v{S}\fi{}imkovic}}, \bibinfo {author}
  {\bibfnamefont {A.}~\bibnamefont {Smetana}},\ and\ \bibinfo {author}
  {\bibfnamefont {P.}~\bibnamefont {Vogel}},\ }\bibfield  {title} {\bibinfo
  {title} {$0\ensuremath{\nu}\ensuremath{\beta}\ensuremath{\beta}$ and
  $2\ensuremath{\nu}\ensuremath{\beta}\ensuremath{\beta}$ nuclear matrix
  elements evaluated in closure approximation, neutrino potentials and su(4)
  symmetry},\ }\href {https://doi.org/10.1103/PhysRevC.98.064325} {\bibfield
  {journal} {\bibinfo  {journal} {Phys. Rev. C}\ }\textbf {\bibinfo {volume}
  {98}},\ \bibinfo {pages} {064325} (\bibinfo {year} {2018})}\BibitemShut
  {NoStop}%
\bibitem [{\citenamefont {Lv}\ \emph {et~al.}(2023)\citenamefont {Lv},
  \citenamefont {Niu}, \citenamefont {Fang}, \citenamefont {Yao}, \citenamefont
  {Bai},\ and\ \citenamefont {Meng}}]{Lv2023Phys.Rev.CL051304}%
  \BibitemOpen
  \bibfield  {author} {\bibinfo {author} {\bibfnamefont {W.-L.}\ \bibnamefont
  {Lv}}, \bibinfo {author} {\bibfnamefont {Y.-F.}\ \bibnamefont {Niu}},
  \bibinfo {author} {\bibfnamefont {D.-L.}\ \bibnamefont {Fang}}, \bibinfo
  {author} {\bibfnamefont {J.-M.}\ \bibnamefont {Yao}}, \bibinfo {author}
  {\bibfnamefont {C.-L.}\ \bibnamefont {Bai}},\ and\ \bibinfo {author}
  {\bibfnamefont {J.}~\bibnamefont {Meng}},\ }\bibfield  {title} {\bibinfo
  {title} {$0\ensuremath{\nu}\ensuremath{\beta}\ensuremath{\beta}$-decay
  nuclear matrix elements in self-consistent skyrme quasiparticle random-phase
  approximation: Uncertainty from pairing interaction},\ }\href
  {https://doi.org/10.1103/PhysRevC.108.L051304} {\bibfield  {journal}
  {\bibinfo  {journal} {Phys. Rev. C}\ }\textbf {\bibinfo {volume} {108}},\
  \bibinfo {pages} {L051304} (\bibinfo {year} {2023})}\BibitemShut {NoStop}%
\bibitem [{\citenamefont {Jokiniemi}\ and\ \citenamefont
  {Men\'endez}(2023)}]{Jokiniemi2023Phys.Rev.C044316}%
  \BibitemOpen
  \bibfield  {author} {\bibinfo {author} {\bibfnamefont {L.}~\bibnamefont
  {Jokiniemi}}\ and\ \bibinfo {author} {\bibfnamefont {J.}~\bibnamefont
  {Men\'endez}},\ }\bibfield  {title} {\bibinfo {title} {Correlations between
  neutrinoless double-$\ensuremath{\beta}$, double gamow-teller, and
  double-magnetic decays in the proton-neutron quasiparticle random-phase
  approximation framework},\ }\href
  {https://doi.org/10.1103/PhysRevC.107.044316} {\bibfield  {journal} {\bibinfo
   {journal} {Phys. Rev. C}\ }\textbf {\bibinfo {volume} {107}},\ \bibinfo
  {pages} {044316} (\bibinfo {year} {2023})}\BibitemShut {NoStop}%
\bibitem [{\citenamefont {Yao}\ \emph {et~al.}(2022{\natexlab{b}})\citenamefont
  {Yao}, \citenamefont {Ginnett}, \citenamefont {Belley}, \citenamefont
  {Miyagi}, \citenamefont {Wirth}, \citenamefont {Bogner}, \citenamefont
  {Engel}, \citenamefont {Hergert}, \citenamefont {Holt},\ and\ \citenamefont
  {Stroberg}}]{Yao2022Phys.Rev.C014315}%
  \BibitemOpen
  \bibfield  {author} {\bibinfo {author} {\bibfnamefont {J.~M.}\ \bibnamefont
  {Yao}}, \bibinfo {author} {\bibfnamefont {I.}~\bibnamefont {Ginnett}},
  \bibinfo {author} {\bibfnamefont {A.}~\bibnamefont {Belley}}, \bibinfo
  {author} {\bibfnamefont {T.}~\bibnamefont {Miyagi}}, \bibinfo {author}
  {\bibfnamefont {R.}~\bibnamefont {Wirth}}, \bibinfo {author} {\bibfnamefont
  {S.}~\bibnamefont {Bogner}}, \bibinfo {author} {\bibfnamefont
  {J.}~\bibnamefont {Engel}}, \bibinfo {author} {\bibfnamefont
  {H.}~\bibnamefont {Hergert}}, \bibinfo {author} {\bibfnamefont {J.~D.}\
  \bibnamefont {Holt}},\ and\ \bibinfo {author} {\bibfnamefont {S.~R.}\
  \bibnamefont {Stroberg}},\ }\bibfield  {title} {\bibinfo {title} {Ab initio
  studies of the double--gamow-teller transition and its correlation with
  neutrinoless double-$\ensuremath{\beta}$ decay},\ }\href
  {https://doi.org/10.1103/PhysRevC.106.014315} {\bibfield  {journal} {\bibinfo
   {journal} {Phys. Rev. C}\ }\textbf {\bibinfo {volume} {106}},\ \bibinfo
  {pages} {014315} (\bibinfo {year} {2022}{\natexlab{b}})}\BibitemShut
  {NoStop}%
\bibitem [{\citenamefont {Caurier}\ \emph {et~al.}(2005)\citenamefont
  {Caurier}, \citenamefont {Mart\'{\i}nez-Pinedo}, \citenamefont {Nowacki},
  \citenamefont {Poves},\ and\ \citenamefont
  {Zuker}}]{Caurier2005Rev.Mod.Phys.427488}%
  \BibitemOpen
  \bibfield  {author} {\bibinfo {author} {\bibfnamefont {E.}~\bibnamefont
  {Caurier}}, \bibinfo {author} {\bibfnamefont {G.}~\bibnamefont
  {Mart\'{\i}nez-Pinedo}}, \bibinfo {author} {\bibfnamefont {F.}~\bibnamefont
  {Nowacki}}, \bibinfo {author} {\bibfnamefont {A.}~\bibnamefont {Poves}},\
  and\ \bibinfo {author} {\bibfnamefont {A.~P.}\ \bibnamefont {Zuker}},\
  }\bibfield  {title} {\bibinfo {title} {The shell model as a unified view of
  nuclear structure},\ }\href {https://doi.org/10.1103/RevModPhys.77.427}
  {\bibfield  {journal} {\bibinfo  {journal} {Rev. Mod. Phys.}\ }\textbf
  {\bibinfo {volume} {77}},\ \bibinfo {pages} {427} (\bibinfo {year}
  {2005})}\BibitemShut {NoStop}%
\bibitem [{\citenamefont {Meng}(2016)}]{Meng2016}%
  \BibitemOpen
  \bibinfo {editor} {\bibfnamefont {J.}~\bibnamefont {Meng}},\ ed.,\ \href@noop
  {} {\emph {\bibinfo {title} {Relativistic Density Functional for Nuclear
  Structure}}},\ \bibinfo {series} {International Review of Nuclear Physics},
  Vol.~\bibinfo {volume} {10}\ (\bibinfo  {publisher} {World Scientific,
  Singapore},\ \bibinfo {year} {2016})\BibitemShut {NoStop}%
\bibitem [{\citenamefont {Zhao}\ \emph {et~al.}(2016)\citenamefont {Zhao},
  \citenamefont {Ring},\ and\ \citenamefont {Meng}}]{Zhao2016Phys.Rev.C041301}%
  \BibitemOpen
  \bibfield  {author} {\bibinfo {author} {\bibfnamefont {P.~W.}\ \bibnamefont
  {Zhao}}, \bibinfo {author} {\bibfnamefont {P.}~\bibnamefont {Ring}},\ and\
  \bibinfo {author} {\bibfnamefont {J.}~\bibnamefont {Meng}},\ }\bibfield
  {title} {\bibinfo {title} {Configuration interaction in symmetry-conserving
  covariant density functional theory},\ }\href
  {https://doi.org/10.1103/PhysRevC.94.041301} {\bibfield  {journal} {\bibinfo
  {journal} {Phys. Rev. C}\ }\textbf {\bibinfo {volume} {94}},\ \bibinfo
  {pages} {041301} (\bibinfo {year} {2016})}\BibitemShut {NoStop}%
\bibitem [{\citenamefont {Wang}\ \emph {et~al.}(2022)\citenamefont {Wang},
  \citenamefont {Zhao},\ and\ \citenamefont {Meng}}]{Wang2022Phys.Rev.C054311}%
  \BibitemOpen
  \bibfield  {author} {\bibinfo {author} {\bibfnamefont {Y.~K.}\ \bibnamefont
  {Wang}}, \bibinfo {author} {\bibfnamefont {P.~W.}\ \bibnamefont {Zhao}},\
  and\ \bibinfo {author} {\bibfnamefont {J.}~\bibnamefont {Meng}},\ }\bibfield
  {title} {\bibinfo {title} {Configuration-interaction projected density
  functional theory: Effects of four-quasiparticle configurations and time-odd
  interactions},\ }\href {https://doi.org/10.1103/PhysRevC.105.054311}
  {\bibfield  {journal} {\bibinfo  {journal} {Phys. Rev. C}\ }\textbf {\bibinfo
  {volume} {105}},\ \bibinfo {pages} {054311} (\bibinfo {year}
  {2022})}\BibitemShut {NoStop}%
\bibitem [{\citenamefont {Wang}\ \emph {et~al.}(2024)\citenamefont {Wang},
  \citenamefont {Zhao},\ and\ \citenamefont
  {Meng}}]{Wang2024Phys.Lett.B138346}%
  \BibitemOpen
  \bibfield  {author} {\bibinfo {author} {\bibfnamefont {Y.~K.}\ \bibnamefont
  {Wang}}, \bibinfo {author} {\bibfnamefont {P.~W.}\ \bibnamefont {Zhao}},\
  and\ \bibinfo {author} {\bibfnamefont {J.}~\bibnamefont {Meng}},\ }\bibfield
  {title} {\bibinfo {title} {Nuclear chiral rotation within relativistic
  configuration-interaction density functional theory},\ }\href
  {https://doi.org/https://doi.org/10.1016/j.physletb.2023.138346} {\bibfield
  {journal} {\bibinfo  {journal} {Phys. Lett. B}\ }\textbf {\bibinfo {volume}
  {848}},\ \bibinfo {pages} {138346} (\bibinfo {year} {2024})}\BibitemShut
  {NoStop}%
\bibitem [{\citenamefont {Wang}\ \emph {et~al.}(2023)\citenamefont {Wang},
  \citenamefont {Zhao},\ and\ \citenamefont
  {Meng}}]{Wang2023arXivpreprintarXiv2304.12009}%
  \BibitemOpen
  \bibfield  {author} {\bibinfo {author} {\bibfnamefont {Y.~K.}\ \bibnamefont
  {Wang}}, \bibinfo {author} {\bibfnamefont {P.~W.}\ \bibnamefont {Zhao}},\
  and\ \bibinfo {author} {\bibfnamefont {J.}~\bibnamefont {Meng}},\ }\href@noop
  {} {\bibfield  {journal} {\bibinfo  {journal} {arXiv:2304.12009}\ } (\bibinfo
  {year} {2023})}\BibitemShut {NoStop}%
\bibitem [{\citenamefont {Jiao}\ \emph {et~al.}(2017)\citenamefont {Jiao},
  \citenamefont {Engel},\ and\ \citenamefont
  {Holt}}]{Jiao2017Phys.Rev.C054310}%
  \BibitemOpen
  \bibfield  {author} {\bibinfo {author} {\bibfnamefont {C.~F.}\ \bibnamefont
  {Jiao}}, \bibinfo {author} {\bibfnamefont {J.}~\bibnamefont {Engel}},\ and\
  \bibinfo {author} {\bibfnamefont {J.~D.}\ \bibnamefont {Holt}},\ }\bibfield
  {title} {\bibinfo {title} {Neutrinoless double-$\ensuremath{\beta}$ decay
  matrix elements in large shell-model spaces with the generator-coordinate
  method},\ }\href {https://doi.org/10.1103/PhysRevC.96.054310} {\bibfield
  {journal} {\bibinfo  {journal} {Phys. Rev. C}\ }\textbf {\bibinfo {volume}
  {96}},\ \bibinfo {pages} {054310} (\bibinfo {year} {2017})}\BibitemShut
  {NoStop}%
\bibitem [{\citenamefont {Wang}\ \emph {et~al.}(2021)\citenamefont {Wang},
  \citenamefont {Zhao},\ and\ \citenamefont {Meng}}]{Wang2021Phys.Rev.C014320}%
  \BibitemOpen
  \bibfield  {author} {\bibinfo {author} {\bibfnamefont {Y.~K.}\ \bibnamefont
  {Wang}}, \bibinfo {author} {\bibfnamefont {P.~W.}\ \bibnamefont {Zhao}},\
  and\ \bibinfo {author} {\bibfnamefont {J.}~\bibnamefont {Meng}},\ }\bibfield
  {title} {\bibinfo {title} {Nuclear matrix elements of neutrinoless
  double-$\ensuremath{\beta}$ decay in the triaxial projected shell model},\
  }\href {https://doi.org/10.1103/PhysRevC.104.014320} {\bibfield  {journal}
  {\bibinfo  {journal} {Phys. Rev. C}\ }\textbf {\bibinfo {volume} {104}},\
  \bibinfo {pages} {014320} (\bibinfo {year} {2021})}\BibitemShut {NoStop}%
\bibitem [{\citenamefont {Song}\ \emph {et~al.}(2014)\citenamefont {Song},
  \citenamefont {Yao}, \citenamefont {Ring},\ and\ \citenamefont
  {Meng}}]{Song2014Phys.Rev.C054309}%
  \BibitemOpen
  \bibfield  {author} {\bibinfo {author} {\bibfnamefont {L.~S.}\ \bibnamefont
  {Song}}, \bibinfo {author} {\bibfnamefont {J.~M.}\ \bibnamefont {Yao}},
  \bibinfo {author} {\bibfnamefont {P.}~\bibnamefont {Ring}},\ and\ \bibinfo
  {author} {\bibfnamefont {J.}~\bibnamefont {Meng}},\ }\bibfield  {title}
  {\bibinfo {title} {Relativistic description of nuclear matrix elements in
  neutrinoless double-$\ensuremath{\beta}$ decay},\ }\href
  {https://doi.org/10.1103/PhysRevC.90.054309} {\bibfield  {journal} {\bibinfo
  {journal} {Phys. Rev. C}\ }\textbf {\bibinfo {volume} {90}},\ \bibinfo
  {pages} {054309} (\bibinfo {year} {2014})}\BibitemShut {NoStop}%
\bibitem [{\citenamefont {Sen'kov}\ and\ \citenamefont
  {Horoi}(2013)}]{Senkov2013Phys.Rev.C064312}%
  \BibitemOpen
  \bibfield  {author} {\bibinfo {author} {\bibfnamefont {R.~A.}\ \bibnamefont
  {Sen'kov}}\ and\ \bibinfo {author} {\bibfnamefont {M.}~\bibnamefont
  {Horoi}},\ }\bibfield  {title} {\bibinfo {title} {Neutrinoless
  double-$\ensuremath{\beta}$ decay of ${}^{48}$ca in the shell model: Closure
  versus nonclosure approximation},\ }\href
  {https://doi.org/10.1103/PhysRevC.88.064312} {\bibfield  {journal} {\bibinfo
  {journal} {Phys. Rev. C}\ }\textbf {\bibinfo {volume} {88}},\ \bibinfo
  {pages} {064312} (\bibinfo {year} {2013})}\BibitemShut {NoStop}%
\bibitem [{\citenamefont {Ring}\ and\ \citenamefont {Schuck}(2004)}]{Ring2004}%
  \BibitemOpen
  \bibfield  {author} {\bibinfo {author} {\bibfnamefont {P.}~\bibnamefont
  {Ring}}\ and\ \bibinfo {author} {\bibfnamefont {P.}~\bibnamefont {Schuck}},\
  }\href@noop {} {\emph {\bibinfo {title} {The nuclear many-body problem}}}\
  (\bibinfo  {publisher} {Springer Science \& Business Media, New York},\
  \bibinfo {year} {2004})\BibitemShut {NoStop}%
\bibitem [{\citenamefont {Carlsson}\ and\ \citenamefont
  {Rotureau}(2021)}]{Carlsson2021Phys.Rev.Lett.172501}%
  \BibitemOpen
  \bibfield  {author} {\bibinfo {author} {\bibfnamefont {B.~G.}\ \bibnamefont
  {Carlsson}}\ and\ \bibinfo {author} {\bibfnamefont {J.}~\bibnamefont
  {Rotureau}},\ }\bibfield  {title} {\bibinfo {title} {New and practical
  formulation for overlaps of bogoliubov vacua},\ }\href
  {https://doi.org/10.1103/PhysRevLett.126.172501} {\bibfield  {journal}
  {\bibinfo  {journal} {Phys. Rev. Lett.}\ }\textbf {\bibinfo {volume} {126}},\
  \bibinfo {pages} {172501} (\bibinfo {year} {2021})}\BibitemShut {NoStop}%
\bibitem [{\citenamefont {Hu}\ \emph {et~al.}(2014)\citenamefont {Hu},
  \citenamefont {Gao},\ and\ \citenamefont
  {Chen}}]{Hu2014PhysicsLettersB162166}%
  \BibitemOpen
  \bibfield  {author} {\bibinfo {author} {\bibfnamefont {Q.~L.}\ \bibnamefont
  {Hu}}, \bibinfo {author} {\bibfnamefont {Z.~C.}\ \bibnamefont {Gao}},\ and\
  \bibinfo {author} {\bibfnamefont {Y.~S.}\ \bibnamefont {Chen}},\ }\bibfield
  {title} {\bibinfo {title} {Matrix elements of one-body and two-body operators
  between arbitrary hfb multi-quasiparticle states},\ }\href@noop {} {\bibfield
   {journal} {\bibinfo  {journal} {Phys. Lett. B}\ }\textbf {\bibinfo {volume}
  {734}},\ \bibinfo {pages} {162} (\bibinfo {year} {2014})}\BibitemShut
  {NoStop}%
\bibitem [{\citenamefont {Zhao}\ \emph {et~al.}(2010)\citenamefont {Zhao},
  \citenamefont {Li}, \citenamefont {Yao},\ and\ \citenamefont
  {Meng}}]{Zhao2010Phys.Rev.C054319}%
  \BibitemOpen
  \bibfield  {author} {\bibinfo {author} {\bibfnamefont {P.~W.}\ \bibnamefont
  {Zhao}}, \bibinfo {author} {\bibfnamefont {Z.~P.}\ \bibnamefont {Li}},
  \bibinfo {author} {\bibfnamefont {J.~M.}\ \bibnamefont {Yao}},\ and\ \bibinfo
  {author} {\bibfnamefont {J.}~\bibnamefont {Meng}},\ }\bibfield  {title}
  {\bibinfo {title} {New parametrization for the nuclear covariant energy
  density functional with a point-coupling interaction},\ }\href
  {https://doi.org/10.1103/PhysRevC.82.054319} {\bibfield  {journal} {\bibinfo
  {journal} {Phys. Rev. C}\ }\textbf {\bibinfo {volume} {82}},\ \bibinfo
  {pages} {054319} (\bibinfo {year} {2010})}\BibitemShut {NoStop}%
\bibitem [{\citenamefont {B\"urvenich}\ \emph {et~al.}(2002)\citenamefont
  {B\"urvenich}, \citenamefont {Madland}, \citenamefont {Maruhn},\ and\
  \citenamefont {Reinhard}}]{Buervenich2002Phys.Rev.C044308}%
  \BibitemOpen
  \bibfield  {author} {\bibinfo {author} {\bibfnamefont {T.}~\bibnamefont
  {B\"urvenich}}, \bibinfo {author} {\bibfnamefont {D.~G.}\ \bibnamefont
  {Madland}}, \bibinfo {author} {\bibfnamefont {J.~A.}\ \bibnamefont
  {Maruhn}},\ and\ \bibinfo {author} {\bibfnamefont {P.-G.}\ \bibnamefont
  {Reinhard}},\ }\bibfield  {title} {\bibinfo {title} {Nuclear ground state
  observables and qcd scaling in a refined relativistic point coupling model},\
  }\href {https://doi.org/10.1103/PhysRevC.65.044308} {\bibfield  {journal}
  {\bibinfo  {journal} {Phys. Rev. C}\ }\textbf {\bibinfo {volume} {65}},\
  \bibinfo {pages} {044308} (\bibinfo {year} {2002})}\BibitemShut {NoStop}%
\bibitem [{\citenamefont {Tian}\ \emph {et~al.}(2009)\citenamefont {Tian},
  \citenamefont {Ma},\ and\ \citenamefont
  {Ring}}]{Tian2009PhysicsLettersB4450}%
  \BibitemOpen
  \bibfield  {author} {\bibinfo {author} {\bibfnamefont {Y.}~\bibnamefont
  {Tian}}, \bibinfo {author} {\bibfnamefont {Z.~Y.}\ \bibnamefont {Ma}},\ and\
  \bibinfo {author} {\bibfnamefont {P.}~\bibnamefont {Ring}},\ }\bibfield
  {title} {\bibinfo {title} {A finite range pairing force for density
  functional theory in superfluid nuclei},\ }\href
  {https://doi.org/https://doi.org/10.1016/j.physletb.2009.04.067} {\bibfield
  {journal} {\bibinfo  {journal} {Phys. Lett. B}\ }\textbf {\bibinfo {volume}
  {676}},\ \bibinfo {pages} {44} (\bibinfo {year} {2009})}\BibitemShut
  {NoStop}%
\bibitem [{\citenamefont {Yao}\ \emph {et~al.}(2015)\citenamefont {Yao},
  \citenamefont {Song}, \citenamefont {Hagino}, \citenamefont {Ring},\ and\
  \citenamefont {Meng}}]{Yao2015Phys.Rev.C024316}%
  \BibitemOpen
  \bibfield  {author} {\bibinfo {author} {\bibfnamefont {J.~M.}\ \bibnamefont
  {Yao}}, \bibinfo {author} {\bibfnamefont {L.~S.}\ \bibnamefont {Song}},
  \bibinfo {author} {\bibfnamefont {K.}~\bibnamefont {Hagino}}, \bibinfo
  {author} {\bibfnamefont {P.}~\bibnamefont {Ring}},\ and\ \bibinfo {author}
  {\bibfnamefont {J.}~\bibnamefont {Meng}},\ }\bibfield  {title} {\bibinfo
  {title} {Systematic study of nuclear matrix elements in neutrinoless
  double-$\ensuremath{\beta}$ decay with a beyond-mean-field covariant density
  functional theory},\ }\href {https://doi.org/10.1103/PhysRevC.91.024316}
  {\bibfield  {journal} {\bibinfo  {journal} {Phys. Rev. C}\ }\textbf {\bibinfo
  {volume} {91}},\ \bibinfo {pages} {024316} (\bibinfo {year}
  {2015})}\BibitemShut {NoStop}%
\end{thebibliography}
%

\end{document}